\documentclass[prl,aps,amssymb,showpacs,twocolumn]{revtex4}
\usepackage{amsmath}
\usepackage{amssymb}
\usepackage{amsthm}
\usepackage{amsfonts}
\usepackage{enumerate}
\usepackage{latexsym}
\usepackage{bm}
\usepackage{graphicx}
\usepackage{subfigure}

\newcommand{\beq}{\begin{equation}}
\newcommand{\eneq}{\end{equation}}

\input{epsf}

\begin{document}

\tolerance 10000

\newcommand{\vk}{{\bf k}}


\title{The Anatomy of Abelian and Non-Abelian Fractional Quantum Hall States}

\author{B. Andrei Bernevig$^{1,2}$ and N. Regnault$^{3}$}

\affiliation{$^1$ Department of Physics, Princeton University, Princeton, NJ 08544}
\affiliation{$^2$ Princeton Center for Theoretical Science, Princeton, NJ 08544}
\affiliation{$^3$ Laboratoire Pierre Aigrain, Departement de Physique, ENS, CNRS, 24 rue Lhomond, 75005 Paris,  France}

\begin{abstract}
 We deduce a new set of symmetries and relations between the coefficients of the expansion of Abelian and Non-Abelian Fractional Quantum Hall (FQH) states in free (bosonic or fermionic) many-body states. Our rules allow to build an approximation of a FQH model state with an overlap increasing with growing system size (that may sometimes reach unity!) while using a fraction of the original Hilbert space. We prove these symmetries by deriving a previously unknown recursion formula for all the coefficients of the Slater expansion of the Laughlin, Read Rezayi and many other states (all Jacks multiplied by Vandermonde determinants), which completely removes the current need for diagonalization procedures.
\end{abstract}

\date{\today}

\pacs{73.43.–f, 11.25.Hf}

\maketitle

Model wavefunctions, such as the Laughlin \cite{laughlin1983}, Moore-Read (MR) \cite{moore1991}, and Read-Rezayi (RR) \cite{read1999} states, have so far provided the key to
understand the physics of Abelian and non-Abelian FQH phases. Despite their explicit availability (in some cases) in terms of the electron positions, their expansions in the
non-interacting basis of occupation number states (Slater determinants or monomials) is unknown and is considered intractable. As a result, quantitative studies of these states have heavily relied on exact diagonalization methods \cite{haldane1985}.

In usual numerical methods, one starts from a model \cite{haldane1985} Hamiltonian, generates the Lowest Landau Level (LLL) Hilbert space and diagonalizes the Hamiltonian within this space. However, one immediately hits an insurmountable barrier: diagonalizing a Hamiltonian matrix of a factorially growing size. The only known symmetry of the coefficients of the expansion of a FQH state in Slater determinants is that of the angular momentum $L_z \rightarrow - L_z$: on the sphere this reflects the indistinguishability of the North and South poles. This symmetry, only valid when the $L_z=0$ (useless for quasihole excitations), roughly halves the size of the Hilbert space needed to construct the model FQH state. It would hence be very beneficial to discover other symmetry rules that the expansion coefficients satisfy. There exist several past attempts at identifying the coefficients of the free many body states in the simplest interacting Laughlin $1/3$ state \cite{difrancesco1994,dunne1993}. However, those attempts can only obtain a small number  $O(1/N!)$ of these coefficients, a statement which becomes painfully obvious when the maximum size Laughlin state generated by these methods is still much less than what can be achieved by exact diagonalization studies.

\begin{figure}
\includegraphics[width=3.3in, height=1.8in]{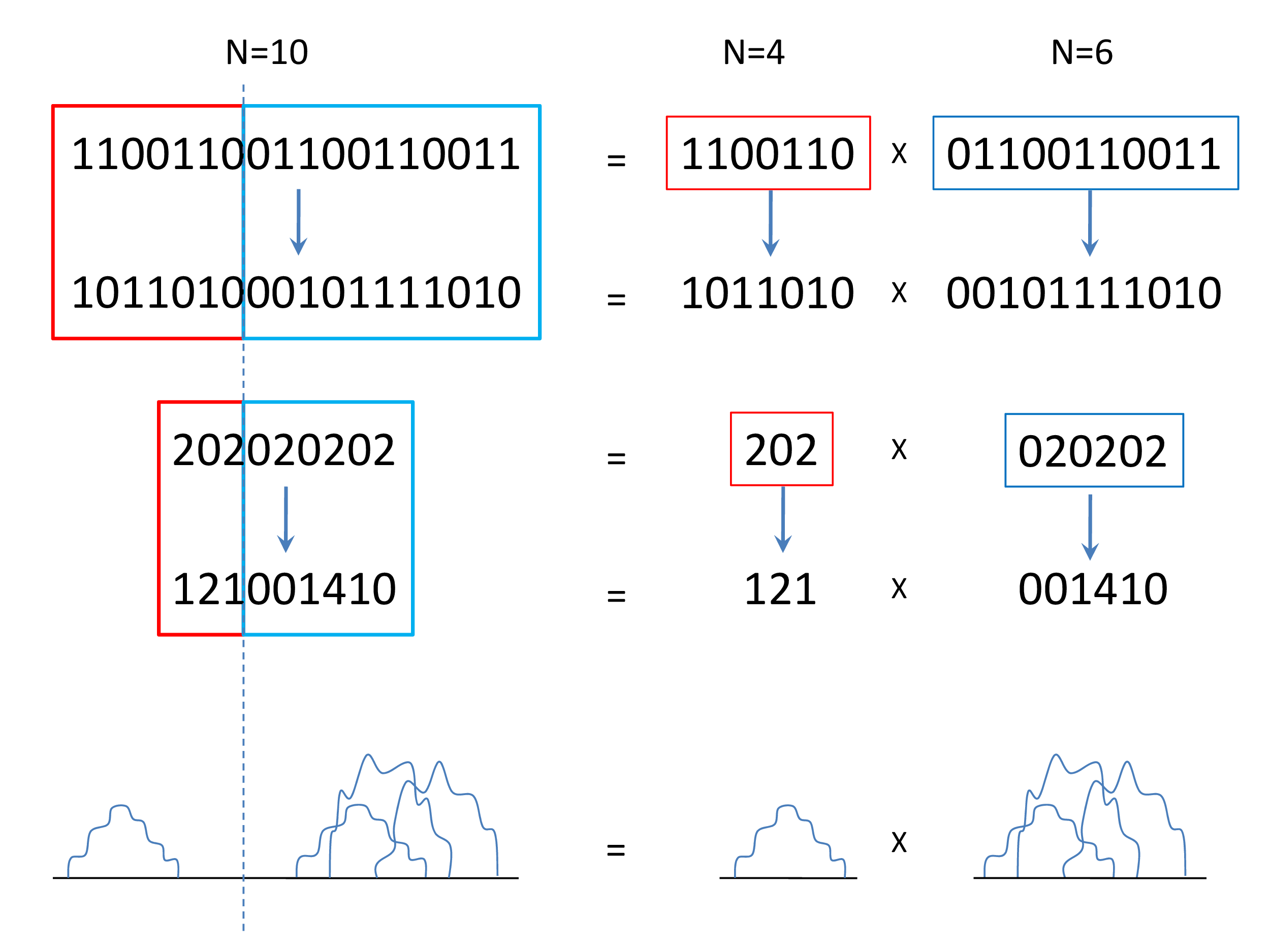}
\caption{Example of the product rule (see text) for fermionic and bosonic MR States. The rule is similar in idea to the computation of Feynman disconnected diagrams.}\label{productrule}
\end{figure}

In the present letter we discover a new series of rules that the coefficients of model FQH states satisfy. Our rules are valid for all bosonic and fermionic Jack polynomial states which include the Laughlin, MR, and RR series. We first obtain an explicit recurrence relation for the coefficients of a fermionic model FQH state that completely removes the need to use diagonalization techniques, hence removing the most important obstacle to achieving larger size model wavefunctions. We then use this to obtain several rules that the coefficients of the expansion satisfy. Some of these symmetries are reminiscent of product rules in Feynman disconnected diagrams. One specific case of our rules is equivalent to the existence of perfect off-diagonal long range order (ODLRO); other cases explain small parts of the observed entanglement spectrum (ES) \cite{li2008}. For Laughlin and MR, the application of these rules roughly halves the Hilbert space needed to produce these states. Still the overlap with the exact state is increasing with system size and can approach unity in the thermodynamic limit. We implement our new methods to present a proof of principle break of the current size barrier and obtain the MR state for $24$ particles, with Hilbert space $\approx 20^4$ times larger than current exact diagonalization techniques. A future paper will use the new muscle to compute quasihole propagators and topological properties.

All the FQH states discussed in this paper are squeezed polynomials: in the development onto occupation number
(Fock basis) states (monomials for bosons and slaters for fermions), we only find orbital occupations obtained by a
so-called squeezing operation on the root occupation configuration \cite{haldaneUCSB2006,bernevig2007}. We represent an angular momentum partition $\lambda$ with length $\ell_{\lambda} \le
N$ as a occupation-number configuration $n(\lambda)$ =
$\{n_m(\lambda),m=0,1,2,\ldots\}$ of each of the LLL orbitals $\phi_m(z) = (2\pi m! 2^m)^{-1/2} z^m \exp(-|z|^2/4)$ with angular momentum $L_z = m \hbar$, where, for $m > 0$, $n_m(\lambda)$ is the
multiplicity of $m$ in $\lambda$. It can be only $0$ or $1$ for fermionic FQH states, but can be any positive integer for bosonic ones. The formalism used here is thoroughly presented in \cite{haldaneUCSB2006,bernevig2007}. It has been showed \cite{bernevig2007,haldaneUCSB2006} that the $N$-particle bosonic Read-Rezayi series of states (which include Laughlin and MR) are the $r=2$ \emph{single} Jack polynomials (Jacks) $J^\alpha_\lambda(z_1,...,z_N)$ of parameter $\alpha = - \frac{k+1}{r-1}$ and root partition $n(\lambda) = [k0^{r-1}k...k0^{r-1}k]$. Quasihole states can be written as coherent state superpositions of Jacks \cite{bernevig2009}. This identification proves their squeezing property and allows; unfortunately, since the Jacks are symmetric polynomials, only bosonic FQH states can be accessed through this procedure, while the fermionic states are the physically relevant ones. The problem is more severe: the Jacks are expanded in quantum-mechanically orthogonal free-boson many-body states
(monomials) $m_\lambda (z_1,...,z_N) = Per(z_i^{\lambda_j})/\prod_m n_m(\lambda)!$, where $Per$ is the permanent. When multiplied by a Vandermonde determinant, the monomial does not become a Slater determinant, this being the property of the Schur functions. As such, to obtain the Slater decomposition of the fermionic FQH state $J_\lambda^\alpha (z_1,...,z_N)\prod_{i<j} (z_i - z_j)$ one would first have to transform the Jack from monomial to Schur basis. This transformation involves knowledge of all the Kostka numbers, a long-standing unsolved mathematical problem with no known efficient algorithm. We depart from this hopeless route and attack the problem differently: we want to find the coefficients $b_{\lambda \mu}$ of the decomposition:
\begin{equation}
S_\lambda^\alpha(z_1...z_N)= J_{\lambda^B}^\alpha (z_1...z_N)\prod_{i<j}^N (z_i - z_j) = \sum_{\mu \le \lambda} b_{\lambda \mu} \text{sl}_\mu \label{fermionicjack}
\end{equation}
\noindent $\text{sl}_\lambda = Det(z_i^{\lambda_j})$ is the quantum mechanically (unnormalized) orthogonal Slater determinant; all partitions $\mu$ are squeezed from the squeezed partition $\lambda$ \cite{bernevig2007,haldaneUCSB2006} related to the Jack bosonic partition $\lambda^B$ by: $\lambda_i = \lambda^B_i + (N-i)$. By rescaling we choose $b_{\lambda \lambda} =1$. $J^\alpha_{\lambda^B}$ is an eigenstate of the Laplace-Beltrami operator $H_{LB}(\alpha)$ with known eigenvalue \cite{stanley1989}. It is then only a matter of (tedious) algebra to find the new operator that diagonalizes $S_\lambda^\alpha(z_1...z_N)$. We call it Fermionic Laplace-Beltrami $H^F_{LB}(\alpha) S_{\lambda}^\alpha(z_1...z_N) = E_\lambda(\alpha)  S_\lambda^\alpha(z_1...z_N)$, with $E_\lambda = \sum_i \lambda_i (\lambda_i - 2(\frac{1}{\alpha}-1) i)  + (\frac{1}{\alpha} -1)((N+1)\sum_i \lambda_i - N(N-1))$:
\begin{widetext}
\begin{equation}\label{fermioniclaplacebeltrami}
H^F_{LB}(\alpha) = H_{\text{K}} +  \frac{1}{2}\left(\frac{1}{\alpha} -1\right) H_{\text{I}} = \sum_i \left(z_i \frac{\partial}{\partial z_i} \right)^2 + \frac{1}{2}\left(\frac{1}{\alpha} -1\right) \sum_{i\ne j} \frac{z_i + z_j}{z_i - z_j}\left(z_i \frac{\partial}{\partial z_i} - z_j \frac{\partial}{\partial z_j}\right) - 2\frac{z_i^2+ z_j^2}{(z_i- z_j)^2}
\end{equation}
\end{widetext}
\noindent We now consider the action of this operator on a Slater $\text{sl}_\mu$. The action of the kinetic term is $\sum_i H_{\text{k}} \text{sl}_\mu = \sum_i \mu_i^2 \text{sl}_\mu$. The action of the remaining "interaction" part is complicated. We sketch the significant step needed to untangle it. As the "interaction" is a two-body operator, it is revealing to consider its action on the two particle Slater $\text{sl}_{\mu= (\mu_1, \mu_2)} = z_1^{\mu_1} z_2^{\mu_2} -z_2^{\mu_1} z_1^{\mu_2}$. Its action on larger size Slaters can be decomposed in pairwise actions (with sign-keeping complications). We find $H_{\text{I}} \text{sl}_{(\mu_1, \mu_2)}$ equals:
\begin{equation}
 (\mu_1 - \mu_2 - 2) \text{sl}_{(\mu_1, \mu_2)} + 2 \sum_{s=1}^{\frac{\mu_1 -\mu_2}{2}} (\mu_1- \mu_2 - 2 s) \text{sl}_{(\mu_1-s, \mu_2+s)} \nonumber
\end{equation}
\noindent
The interaction term acting on a Slater creates all pairwise squeezed configurations of a partition, but its action is different than that of the corresponding Laplace-Beltrami operator on a monomials.  When generalized to any number of particles, by virtue of dealing with fermions, the above action  acquires a $-1$ sign every time that either $\mu_i-s$ or $\mu_j+s$ crosses another partition component $\mu_k$.  Taking the action of $H^F_{LB}(\alpha)$ on $S_\lambda^\alpha$ in Eq(\ref{fermionicjack}), we rescale the summation over $s$ above and use linear independence of Slaters to obtain:
\begin{equation}
b_{\lambda \mu} = \frac{2\left(\frac{1}{\alpha}-1\right)}{\rho^F_\lambda(\alpha) - \rho^F_\mu(\alpha)} \sum_{\theta; \; \mu<\theta \le \lambda} (\mu_i- \mu_j) b_{\lambda \theta} \cdot (-1)^{\text{N}_{\text{sw}}} \label{fermionicrule}
\end{equation}
The sum is over all partition $\theta = (\mu_1, ...,\mu_i+s, ..., \mu_j-s, ...,\mu_N)$ which strictly dominate $\mu=(\mu_1,...,\mu_N)$ and are squeezed from the root $\lambda$. $\text{N}_{\text{sw}}$ is equal to the number of swaps required to order the partition $\theta$. $\rho^F_\lambda(\alpha) = \sum_i \lambda_i(\lambda_i + 2 i (1- 1/\alpha))$. Eq.(\ref{fermionicrule}) differs from a known relation for the expansion of a Jack polynomial in terms of monomials \cite{ha,Dumitriua2007}: besides different $\rho_\lambda^F(\alpha)$, the factor $(\mu_i - \mu_j)$ in the summand of Eq(\ref{fermionicrule}) does not depend on the partition $\theta$ (does not depend on $s$) whereas it does in the bosonic case $(\mu_i - \mu_j+ 2 s)$. For $\alpha=-(k+1)$, Eq.(\ref{fermionicrule}) gives the coefficients of the RR fermionic states.  It completely removes the need for diagonalization, coefficients being computed iteratively starting from the root partition \cite{RonnyComment}. The algorithm's numerical stability is high and has been checked through some coefficients which are explicitly known. Largest system sizes previously obtained using diagonalization can now be done on a single CPU workstation in a matter of a few hours (for the $1/3$ Laughlin state at $N=15$, the new algorithm requires 500 times less CPU time). The main bottleneck is now the Hilbert space storage which may still be huge.

We now give an example and sketch the proof of a simple rule we found the coefficients of a model FQH state to satisfy. As  above, we work in the un-normalized orbital basis of Slaters or monomials; this completely separates the polynomial part of the problem from the scalar-product part, which involves geometry-dependent normalizations and is hence not fundamental. Once the state is obtained in un-normalized basis, it can be trivially transformed in normalized basis depending on whether we want to perform calculations on the sphere, disk, cylinder or any other 0 genus geometry.

The Moore-Read state for $N=10$ particles is a linear combination of Slater determinants squeezed from the root configuration  $n(\lambda) = 110011001100110011$ \cite{haldaneUCSB2006,bernevig2007}. Let us determine the coefficient of e.g $101101000101111010$ (see Fig[\ref{productrule}]). This configuration has the special property that, if cut in two after the first $7$ orbitals, each resulting partition can be squeezed from its own root partition: at the left of the cut, $1011010$ can obtained by squeezing on the root partition $N=4$ MR root occupation $1100110$; at the right of the cut, $00101111010$ can be squeezed from the $N=6$ MR root occupation $01100110011$ (extra fluxes -zeroes- can be neglected). We proved a product rule: the coefficient of $101101000101111010$ in the $N=10$ particle MR state (140) equals the product of the coefficients of the two disconnected $N=4$ and $N=6$ pieces of the partition: $-2$ and $-70$. A similar property occurs in the bosonic MR state (see Fig[\ref{productrule}]), and for any disconnected occupation numbers. The rule also allows further "dissection" of FQH states. It can be trivially generalized to any product of disconnected pieces. This is a direct consequence of the recursive property of the product rule : some coefficients might be computed from those of two smaller systems which in some case, could be deduced from those of two even smaller systems, etc. This rule bears resemblance to the rules of computation of Feynman disconnected diagrams, where the value of a diagram that can be cut in two disconnected parts can be obtained as the product of the values of the two disconnected parts.

This simple rule valid for all Jack polynomials, bosonic and fermionic, at \emph{any} $\alpha$, of \emph{any} partition. We now very roughly sketch the proof for the fermionic Jacks using our newly found formula. For bosonic Jacks, a similar proof can be obtained. Assume we want to determine the coefficient, in an arbitrary fermionic polynomial $S^\alpha_\lambda (z_1,...,z_N)$, of a configuration $\mu \le \lambda$  that can be divided in two disconnected sets. By assumption $\mu = (\mu_A, \mu_B)$ of $N_A$ and $N_B$ particles ($N_A+ N_B=N$), is supposed to be divisible in two independent partitions: $\mu_A$, squeezed from the root partition of $N_A$ particles $\lambda_{A} = (\lambda_{1},...,\lambda_{N_A})$, and $\mu_B$, squeezed from the root partition of $N_B$ particles $\lambda_{B} = (\lambda_{N_A+1},...,\lambda_{N})$. The proof proceeds by induction. Assume the product rule is valid for ALL partitions, of "separable" form $\theta = (\theta_A, \theta_B)$ where $\mu_A< \theta_A \le \lambda_A$ and $\mu_B< \theta_B \le \lambda_B$. The coefficients of $S^\alpha_\lambda$ are given in Eq(\ref{fermionicjack}) as a recursion of coefficients from partitions which dominate $\mu$. Crucially, if $\mu$ has the separable form chosen by us, then the partitions which dominate $\mu$ and which enter Eq(\ref{fermionicjack}) are also separable, of the form $\theta =(\theta_A, \theta_B)$. Hence the sum in Eq(\ref{fermionicjack}) separates in two distinct sums of $ (\mu_i -\mu_j ) b_{\lambda \theta}(\alpha)\cdot (-1)^{\text{N}_{\text{sw}}} $ over the disconnected parts of the partitions $\theta$: $\sum_{\theta; \; \mu<\theta \le \lambda}  = \sum_{\mu_A < \theta_A \le \lambda_A}  +\sum_{\mu_B < \theta_B \le \lambda_B} $. In the first sum $\mu_i, \mu_j$ belong only to the left hand side $\theta_A$ of the partition and not to $\mu_B$: the partition in the first sum reads $(\theta_A, \mu_B)$. In the second sum they belong only to the right hand side $\theta_B$ of the partition, not to $\mu_A$:the partition in the second sum reads $(\mu_A, \theta_B)$. By the induction assumption, all partitions which dominate $\mu$ satisfy the product rule: $ b_{\lambda (\theta_A, \mu_B) }(\alpha) = b_{\lambda (\theta_A, \lambda_B) }(\alpha)\cdot b_{\lambda (\lambda_A, \mu_B) }(\alpha)$ where $(\theta_A, \lambda_B)$ is denoted as being the partition formed by $\theta_A$ and the ground-state partition $\lambda_B$, and similarly for the other coefficient. The sums then become $\sum_{\mu_A < \theta_A \le \lambda_A}  (\mu_i -\mu_j ) b_{\lambda (\theta_A, \mu_B)}(\alpha)\cdot (-1)^{\text{N}_{\text{sw}}} = b_{\lambda (\lambda_A, \mu_B)}(\alpha) \cdot \sum_{\mu_A < \theta_A \le \lambda_A} (\mu_i -\mu_j ) b_{\lambda (\theta_A, \lambda_B) }(\alpha) \cdot (-1)^{\text{N}_{\text{sw}}}$ and similar for the sum over $\theta_B$. We notice that the sum is proportional to $b_{\lambda (\mu_A, \lambda_B)}$. By using:
 \begin{equation}
\rho_{\lambda}^F(\alpha) - \rho_{\mu}^F(\alpha) = \rho_{\lambda}^F(\alpha)- \rho^F_{\mu_A \lambda_{B}} (\alpha) + \rho^F_{\lambda}(\alpha)- \rho^F_{\lambda_{A}\mu_B} (\alpha) \nonumber
\end{equation}
we can prove that $ b_{\lambda \mu}(\alpha) = b_{\lambda (\mu_A, \lambda_B)}(\alpha)\cdot  b_{\lambda (\lambda_A,\mu_B)}(\alpha)$, which is the product rule described. It is valid for all Jacks, at any $\alpha$ and hence for all Read Rezayi states. We conjecture it is also valid for some non-Jack states \cite{inpreparation}.

 In special cases, we can identify the product rule with physical properties of the FQH states. Cut the $N$-particle Laughlin $\nu=1/2$ state $101010101...101$ into two pieces, one of $2$ orbitals and $1$ particle ($10$) and the second of $N-1$ particles $1010101...101$. The product rule shows that the coefficients of any configuration in the $N$-particle state that has $1$ particle in the first orbital has the same coefficient as the one that the configuration obtained by deleting $10$ from the $N$-particle state would have in the $N-1$ particle state. Since deleting $10$ from the $N$-particle state is equivalent to applying the operator $h^\dagger(z) \psi(z)$ where $h(z) = \prod_{i}^N (z- z_i)^r$ , in this specific case the product rule can be re-written $|\psi_{N-1}\rangle = h^\dagger(z) \psi(z) |\psi_N\rangle$, which is ODLRO statement. In the non-abelian case, a similar statement occurs, when the ODLRO concept is generalized to non-abelian states \cite{inpreparation}.

The product rule also explains aspects of the ES of the FQH states. On the sphere, we cut the state into two hemisphere blocks $A$ and $B$. Following \cite{li2008}, we introduce the ES $\xi$ as $\lambda_i = exp(-\xi_i)$, where $\lambda_i$ are the eigenvalues
of the reduced density matrix $\rho_A$ of one hemisphere. The
eigenvalues can be classified by the number of fermions
$N_A$ in the A block, and also by the total "angular momentum"
$L_z^{(A)}$ of the A block.
It was empirically found for the MR\cite{li2008}, Laughlin \cite{zozulya2009}, and Jain Composite Fermion \cite{regnaultbernevig2009} states that  the low-lying spectrum $\xi_i$ of the reduced density matrix for fixed $N_A$, plotted as a function of $L^{(A)}_z$, displays a structure reflecting the edge CFT. In particular, at maximum $L^{(A)}_z$ one finds one single eigenvalue of the ES, irrespective of where in orbital space the cut was performed. A generic state is expected to have an order $\text{min}(N_A!, (N-N_A)!)$ levels. Our product rule proves that only one eigenvalue is possible at the maximum $L^{(A)}_z$. As we consider the maximum $L^{(A)}_z$, the density matrix at this angular momentum accesses states that are squeezed separately within each hemisphere; if a state would have a configurations squeezed across the cut, then the angular momentum $L^{(A)}_z$ would no longer be maximum. As the states are squeezed separately within each hemisphere, the product rule renders the rows and columns of the density matrix linearly dependent, and hence $1$ non-zero eigenvalue of the Schmidt spectrum.

Given its generality, one can ask whether the product rule works for any $\vec{L}=0$ state in the LLL. It does not. A counter example is the linear superposition of the two states obtained by applying the $\vec{L}=0$ condition on the subspace of states squeezed from the Jain state root partition $2010110102$ \cite{regnaultbernevig2009}. Cutting the state symmetrically in half after $5$ orbitals, the ES generically has $2$ eigenvalues at the largest  $L^{(A)}_z=12$, and the product rule in general doesn't apply. However, since the counting in the ES is \emph{conjectured} to be equivalent to the counting of edge-modes, and since the Jain state has $1$ excitation at maximum $L^{(A)}_z$ - which we showed is equivalent to the existence of a product rule - the rule should work for the Jain state as well in the thermodynamic limit.

 How much of a state can be obtained by the product rule? Assume that the $N$ particle Laughlin state $J^{-2}_{10101...10101} (z_1,...,z_{N})$ is known (this trivially implies that the Laughlin states for smaller number of particles are known). In the $N+1$ Laughlin state $J^{-2}_{1010101...10101} (z_1,...,z_{N+1})$ the configurations that \emph{cannot} be obtained using the product rule are dominated by the configuration where the $2$ particles in the North Pole and South Pole orbital are squeezed together once: $[0110101...1010110]$. Hence, the number of configurations that one can obtain in the $1/2$ Laughlin state is roughly 45-50\% in the thermodynamic limit. For non-abelian states, we can obtain a smaller percentage - for MR, we get roughly 30\% in the thermodynamic limit. We then compute the overlap of the exact $N$-particle Laughlin state with the state obtained by applying our product rule on the exact $N-1$ Laughlin state. We obtain  $0.9875$ for $N=6$ up to $0.9977$ for $N=15$; note that the overlap grows, and likely reaches $1$ in the thermodynamic limit, even though we miss half the configurations of the FQH state. Similarly for the Moore-Read state, the product rule \emph{cannot} obtain configurations squeezed from $[112020202...2020211]$. The overlaps are $0.8858$ for $N=8$ up to $0.9383$ for $N=22$. For fermionic states, the overlaps are a bit lower (like $0.9603$ for  the $1/3$ Laughlin state at $N=15$) but with smaller fraction of the Hilbert space and still increase with system size. We have checked that the product state entanglement entropy is less than $5\%$ higher compared to the model state.

In fact, we can do better. We have also proved a series of rules for configurations that \emph{cannot} be obtained by the product rule.
For example, for Laughlin $\nu=1/2$, the product rule is useless in obtaining the coefficients of any of the configurations squeezed from $01101010...1010110$. However, we have proved another rule that can obtain some of these coefficients. We can prove that the coefficient of the partition $0110\lambda$ in the $N+1$ particle state (where $\lambda$ is any partition squeezed from $1010...1010110$) is proportional to the coefficient of the partition $ 01\lambda$ in the $N$ particle state. The proportionality constant can be obtained by considering the first two members of these partitions $0110101....010110$ and $100110101...010110$. For Laughlin $1/2$ state, the proportionality constant is $-1$. The same rule works for non-abelian states. For example, for the MR state, the product rule is useless in obtaining configurations squeezed from $1120202...2020211$; our new rule proves that the coefficient of the configuration $1120 \lambda$ (where $\lambda$ is squeezed from $202...2020211$) in the $N+2$-particle state is proportional to the coefficient of the configuration $11\lambda$ in the $N$-particle state.  For MR $N=22$, using this rule in addition to the product rules improves the overlap to $0.9417$ while adding only $1.2\%$ of the vector components. The rule above is part of a unified rule that allows the determination of coefficients of all configurations where the particles in the North and South pole have been squeezed only once; using the two rules, for Laughlin $\nu=1/2$ we can obtain all the coefficients for all configurations except the ones squeezed from $0020101..1010200$; for MR, we can obtain all coefficients except the ones squeezed from $0220202...02020220$. The new rule explains the existence of only $1$ eigenvalue in the ES at $L^{(A)}_z = L_{max}-1$, and, unlike the product rule, is only valid for FQH states and their quasiholes (not for all Jacks of arbitrary $\alpha$). Due to lack of space, we will present this rule, as well as other conjectured rules in a separate long publication.

\emph{Conclusion} We have obtained a new formula for the coefficients of the expansion of fermionic model FQH states in Slater determinants that completely removes the need for matrix diagonalization. Using this formula, we have proved a series of new rules of these coefficients that allow further massive reductions in the Hilbert space size necessary to build a state. We hope the rules presented here will redefine the way numerics on model FQH states is currently done.

B.A.B. wishes to thank S.Simon for a critical reading of the manuscript, R. Thomale for pointing out a degeneracy situation in Eq(\ref{fermionicrule}) and especially F.D.M Haldane for countless discussions.

\end{document}